 \documentclass[12pt]{article}
 \usepackage{epsfig}
 \newcommand{\be}{\begin{equation}}
 \newcommand{\ee}{\end{equation}}
 \newcommand{\ba}{\begin{eqnarray}}
 \newcommand{\ea}{\end{eqnarray}}

\def\laq{\raise 0.4ex\hbox{$<$}\kern -0.8em\lower 0.62
ex\hbox{$\sim$}}
\def\gaq{\raise 0.4ex\hbox{$>$}\kern -0.7em\lower 0.62
ex\hbox{$\sim$}}
 \def\Journal#1#2#3#4{{#1} {\bf #2}, #3 (#4)}

 \def\ANP{\em Ann.\ Physics (N.Y.)}
 \def\CQG{\em Class.\ Quantum Grav.}
 
 \def\IJMPA{{\em Int.\ J.\ Mod.\ Phys.} A}
 
 \def\MPLA{{\em Mod.\ Phys.\ Lett.} A}
 \def\NPB{{\em Nucl. Phys.} B}
 
 \def\PLB{{\em Phys. Lett.}  B}
 
 \def\PRD{{\em Phys. Rev.} D}
 \def\TMP{\em Theor.\ Math.\ Phys.}
\begin{document}
\title{Quantum Gravity Corrections to the Schwarzschild Mass}
\author{Marco Cavagli\`a$^{1,2\,}$\thanks{E-mail: cavaglia@mercury.ubi.pt} ~and
Carlo Ungarelli$^1\,$\thanks{Present address: School of Computer Science and
Mathematics, Mercantile House, Hampshire Terrace,
University of Portsmouth, Portsmouth, PO1 2EG, United Kingdom; 
E-mail: ungarell@sun1.sms.port.ac.uk, carlo.ungarelli@port.ac.uk}\\\\
\it
$^1$ Max-Planck-Institut f\"ur Gravitationsphysik,
Albert-Einstein-Institut\\ \it
Am M{\"u}hlenberg 1,
D-14476 Golm, Germany\\\\
\it
$^2$ Departamento de Fisica, Universidade da Beira Interior\\ \it
R.\ Marqu{\^e}s d'{\'A}vila e Bolama, 6200 Covilh{\~a}, Portugal}
\date{\today }
\maketitle

\begin{abstract}
Vacuum spherically symmetric Einstein gravity in $N\ge 4$ dimensions can be
cast in a two-dimensional conformal nonlinear sigma model form by first
integrating on the $(N-2)$-dimensional (hyper)sphere and then performing a
canonical transformation. The conformal sigma model is described by two fields
which are related to the Arnowitt-Deser-Misner mass and to the radius of the
$(N-2)$-dimensional (hyper)sphere, respectively. By quantizing perturbatively
the theory we estimate the quantum corrections to the ADM mass of a black hole.

\medskip\noindent
Pac(s) numbers: 04.60.-m, 04.70.Dy, 04.62.+v
\end{abstract}

\newpage

\noindent
\section{Introduction}
Classically, a neutral, non-rotating, spherically symmetric black hole in
vacuum is completely identified by the value of its (ADM) mass $M_{ADM}$ (see
e.g.\ \cite{MTW}). Since gravity does not couple to any matter field -- and we
impose {\it ab initio} spherical symmetry -- $M_{ADM}$ is constant and the
geometry possesses one extra Killing vector in addition to the Killing vectors
which are associated to the spherical symmetry (Birkhoff theorem). The general
solution of Einstein equations is the famous Schwarzschild metric. It
describes an {\it eternal black hole}.

Naively, we would expect both properties -- the Birkhoff theorem and
$M_{ADM}=constant$ -- to be broken at quantum level. The validity of the
Birkhoff theorem in the quantum canonical theory of spherically symmetric
gravity has been investigated in Refs.\
\cite{Cavaglia-PRD,Cavaglia-Proc,CDF-PhL}. It has been shown that the Birkhoff
theorem holds at quantum level, i.e.\ the quantum theory of spherically
symmetric gravity in vacuum is a quantum mechanical system with a finite number
of degrees of freedom ({\it Quantum Birkhoff Theorem}) \cite{Cavaglia-PRD}.
Moreover, the Hilbert space of the quantum theory is completely determined by
the eigenstates of the (gauge invariant) mass operator.

The aim of this paper is to explore whether the other classical property
($M_{ADM}=constant$) holds in the quantum gravity regime as well. The result of
our investigation is that quantum gravity corrections to the Schwarzschild mass
appear at the second order in the curvature perturbative expansion. For
instance, quantum fluctuations of the mass of a four-dimensional black hole
are, for distances much greater than the horizon radius, 
\be
\Delta M_{ADM}\,\laq\, m_{pl}\left({l_{pl}\over
R}\right)^{2}\,,\label{fluctuations}
\ee
where $l_{pl}$ and  $m_{pl}$ are the Planck length and the Planck mass,
respectively (Notations: Here and throughout the paper
we use natural units.)

A number of approximations are needed to obtain Eq.\ (\ref{fluctuations}). We
will discuss them in detail in the following sections. Here let us just
emphasize two important points concerning Eq.\ (\ref{fluctuations}). Firstly,
the quantum theory breaks down on the horizon(s) where the coupling constants
of the perturbative expansion diverge. Therefore, Eq.\ (\ref{fluctuations}) is
strictly valid for distances much greater than the Schwarzschild radius of the
black hole. Secondly, quantum fluctuations vanish for large radii, i.e.\ at
large distances from the black hole. In the asymptotic regime the black hole
behaves classically and the mass is constant. Quantum fluctuations of the
Schwarzschild mass due to pure quantum gravity effects become manifest when the
black hole horizon is approached.

Equation (\ref{fluctuations}) is obtained in the context of the nonlinear sigma
model approach to spherically symmetric gravity whose basic ingredients are
described in Ref.\ \cite{Cavaglia-PRD}. Firstly, $N$-dimensional spherically
symmetric gravity is cast in a dilaton gravity form by integrating over the
$(N-2)$ spherical coordinates. Then, by a canonical field redefinition the
action is transformed in a two-dimensional conformal nonlinear sigma model with
a fixed target metric. The new fields are the dilaton and a gauge invariant
field $M$ which is constant on the classical solutions of the field equations
and can be identified with the ADM mass of the black hole.

The new action can be quantized perturbatively by expanding the metric of the
target space in normal Riemann coordinates \cite{AFM}. Since the expansion
parameter is proportional to the curvature of the manifold, the theory is a
free field theory far away from the black hole horizon and for large ADM mass
in Planck units. The perturbative theory turns out to be infrared and
ultraviolet divergent. Infrared divergences are eliminated by the introduction
of an infrared regulator $m$. The theory is regularized by usual dimensional
regularization techniques. The consistency of the procedure is verified {\em a
posteriori} by calculating the one-loop $\beta$-function. The theory becomes
asymptotically free at large energy scales, where the perturbative regime is
valid and the infrared regulator can be neglected.

The amplitude of the quantum fluctuations of the ADM mass at a given order in
the perturbative expansion can be read straightforwardly from the two-point
Green functions of the theory. This is possible because the nonlinear sigma
model fields are the ADM mass and the dilaton. Since the fields have a direct
geometrical meaning any problem related to their interpretation in terms of
physical quantities disappears. 

The outline of the paper is as follows. In the next section we illustrate the
classical theory of $N$-dimensional ($N\ge 4$) spherically symmetric gravity.
We start with the dilaton gravity description and then introduce the nonlinear
sigma model picture. Although a part of this section reviews previous work
(see \cite{Cavaglia-PRD} and references therein), its content is useful to
make the paper self-contained. Sections 3 and 4 are devoted to the classical
expansion in normal Riemann coordinates and to the perturbative quantization
of the theory, respectively. (The evaluation of the relevant Feynman diagrams
is briefly outlined in appendix.) Finally, in Section 5 we state our
conclusions.
\section{Classical theory}
It is well known \cite{DG,LGK,GK,CV} that for spherically symmetric metrics
the $N$-dimensional Einstein-Hilbert action (the Ricci tensor is defined as in
\cite{LL})
\be
S^{(N)}={1\over 16\pi l_{pl}^{N-2}}\int d^Ny\,
\sqrt{-G}\,R^{(N)}(G)\label{EH}
\ee
can be cast, upon integration on the $N-2$ spherical coordinates, in the
dilaton gravity form
\be
S_{DG}=\int d^2x\, \sqrt{-g}\, \left[\phi R^{(2)}(g)-{d~\over
d\phi}\ln[W(\phi)](\nabla\phi)^2+V(\phi)\right]\,,
\label{action-dilaton} 
\ee
where the dilaton field is related to the radius of the $(N-2)$-dimensional
sphere and $W(\phi)$ and $V(\phi)$ are given functions whose form depends on
the $N$-dimensional metric ansatz. (We neglect surface terms as they are
irrelevant for the following discussion. For a detailed discussion about the
role of boundary terms see e.g.\ \cite{Cavaglia-PRD, KRV, Kuchar} and
references therein.)

Theories of the form (\ref{action-dilaton}) admit the existence of the
(gauge invariant) quantity \cite{Cavaglia-PRD,Filippov}
\be
M=N(\phi)-W(\phi)(\nabla\phi)^2\,,~~~~N(\phi)=\int^\phi
d\phi'[W(\phi')V(\phi')]\label{M}
\ee
which is locally conserved, i.e.\ 
\be
\nabla_\mu M=0\,.\label{eq-diff-M}
\ee
Equation (\ref{eq-diff-M}) can be easily proved by differentiating Eq.\
(\ref{M}) and using the field equations
\ba
&&\nabla_{(\mu}\nabla_{\nu)}\phi-g_{\mu\nu}\nabla^2\phi
+{1\over 2}g_{\mu\nu}V(\phi)+\displaystyle{d~\over d\phi}\ln[W(\phi)]
[\nabla_{(\mu}\nabla_{\nu)}\phi-{1\over 2}g_{\mu\nu}(\nabla\phi)^2]=0\,,\\
&&R^{(2)}(g)+2\nabla^2 \ln[W(\phi)]+{dV(\phi)\over d\phi}=0\,.
\ea
A further property of $M$ is conformal (Weyl) invariance \cite{Cadoni}. Indeed,
by rescaling the two-dimensional metric \cite{KKL}
\be
g_{\mu\nu}(x)\to g_{\mu\nu}(x)\, A(\phi)\,,\label{Weyl}
\ee
$V(\phi)$ and $W(\phi)$ transform as
\begin{equation}\begin{array}{lll}
V(\phi)&\to&V(\phi)\,/A(\phi)\,,\\\\
W(\phi)&\to&W(\phi)\,A(\phi)\,.
\label{Weyl-2}
\end{array}\end{equation}
Equation (\ref{M}) is clearly invariant under (\ref{Weyl})-(\ref{Weyl-2}).
Using Eqs.\ (\ref{Weyl})-(\ref{Weyl-2}), the action (\ref{action-dilaton}) can
be cast in a simpler form by a suitable choice of $A(\phi)$. Here and
throughout the paper we will set $A(\phi)=1/W(\phi)$ which corresponds
to choosing the spherically symmetric ansatz ($\alpha,\beta=0,\dots N-1$,
$\mu,\nu=0,1$) \cite{CV} 
\begin{equation}\begin{array}{lll}
&ds^2_N&=G_{\alpha\beta}dy^\alpha dy^\beta\\\\
&&=[\phi(x)]^{-(N-3)/(N-2)}\,g_{\mu\nu}(x)\,dx^\mu dx^\nu +
[\gamma\phi(x)]^{2/(N-2)}\,d\Omega^2_{N-2}\,,~~~~~\phi>0\,.
\label{metric-N}
\end{array}\end{equation}
With this choice $W(\phi)\to 1$ and the dilaton gravity 
action (\ref{action-dilaton})
becomes
\be
S_{DG}=\int d^2x\, \sqrt{-g}\,\left[\phi\, R^{(2)}(g)+V(\phi)
\right]\,,\label{action-dilaton-2} 
\ee
where
\be
V(\phi)=(N-2)(N-3)(\gamma^2\phi)^{-1/(N-2)}\,.  
\label{V}
\ee
Here $\gamma=16\pi \, l_{pl}^{N-2}/V_{N-2}$ and   
$V_{N-2}=2\pi^{(N-1)/2}/\Gamma[(N-1)/2]$ is the volume of the
$(N-2)$-dimensional unit sphere $d\Omega^2_{N-2}$. On the gauge shell the
quantity $M$ coincides, apart from some numerical factors, with the ADM
\cite{MTW} mass 
\be
M_{\rm ADM}={\gamma^{1/(N-2)}\over N-2}M\,.\label{M-ADM}
\ee
This property will be essential in the following.

The dilaton gravity action (\ref{action-dilaton}) can be cast in a nonlinear
conformal sigma model form. Here our treatment follows closely
\cite{Cavaglia-PRD}. In two-dimensions the Ricci scalar $R^{(2)}(g)$ can be
locally written as
\be
R^{(2)}(g)=2\,\nabla_\mu A^\mu\,,~~~~A^\mu=
{\nabla^\mu\nabla^\nu\chi\nabla_\nu\chi-
\nabla_\nu\nabla^\nu\chi\nabla^\mu\chi\over
\nabla_\rho\chi\nabla^\rho\chi}\,,\label{ricci-div}
\ee
where $\chi$ is an arbitrary, non-constant, function of the coordinates. 
Equation (\ref{ricci-div}) can be easily checked using conformal coordinates.
Since Eq.\ (\ref{ricci-div}) is a generally covariant expression, and any
two-dimensional metric can be locally cast in the conformal form by a
coordinate transformation \cite{Eisenhart}, Eq.\ (\ref{ricci-div}) is valid in
any system of coordinates.

Differentiating Eq.\ (\ref{M}), and using Eq.\ (\ref{ricci-div}) with
$\chi=\phi$, the action (\ref{action-dilaton}) can be written as a functional
of $M$ and $\phi$. The result is
\be
S=\int_\Sigma d^2x\,\sqrt{-g}\,{\nabla_\mu\phi\nabla^\mu M
\over N(\phi)-M}+S_{\partial}\,,\label{action-new}
\ee
where $S_{\partial}$ is the surface term
\be
S_{\partial}=2\int_\Sigma 
d^2x\,\sqrt{-g}\,\nabla_\mu\left[\nabla^\mu\phi+\phi
A^\mu\right]\,.\label{surface}
\ee
Let us investigate the classical solutions of Eq.\ (\ref{action-new}).
Varying Eq.\ (\ref{action-new}) with respect to $M$ and $\phi$ we find
\ba
&&\nabla_\mu\nabla^\mu\phi-V(\phi)=0\,,\label{eq-M}\\
&&\nabla_\mu M\nabla^\mu
M+\nabla_\nu\phi\nabla^\nu\phi\nabla_\mu\nabla^\mu M=0\,.\label{eq-phi}
\ea
Equations (\ref{eq-M}) and (\ref{eq-phi}) must be complemented by the
constraints
\be
\nabla_{(\mu}\phi\nabla_{\nu)}M-{1\over
2}g_{\mu\nu}\nabla_\sigma\phi\nabla^\sigma M=0\label{eq-g}
\ee
which are obtained by varying Eq.\ (\ref{action-new}) with respect to the
metric $g_{\mu\nu}$. The general solution of Eqs.\ (\ref{eq-M})-(\ref{eq-g})
can be easily obtained using conformal coordinates. Setting
\be
g_{\mu\nu}=\rho\left(\matrix{0&1\cr
1&0\cr}\right)~~~\to~~~ds^2=2\rho(u,v)\,dudv\,,\label{conf-gauge}
\ee
Eqs.\ (\ref{eq-M})-(\ref{eq-phi}) and the constraints (\ref{eq-g}) read
\begin{equation}\begin{array}{lll}
&&\partial_u\partial_v\phi-\displaystyle{\rho\over 2}V(\phi)\,,\\\\
&&\partial_u\partial_v M+\displaystyle{\partial_u M\partial_v M\over N-M}=0\,,\\\\
&&\partial_u\phi\,\partial_u M=0\,,~~~~~~\partial_v\phi\,\partial_v M=0\,.
\end{array}\label{eq-Mphi}
\end{equation}
>From Eqs.\ (\ref{eq-Mphi}) and (\ref{V}) it follows that $M$ is
constant, $M=M_0$. Using Eq.\ (\ref{eq-Mphi}) and (\ref{M}) the general
solution can be written
\be
M=M_0\,,~~~~~\phi=\phi(\Psi)\,,~~~~~{d\phi\over d\Psi}=N[\phi(\Psi)]-M_0\,,
\ee
where $\Psi=U(u)+V(v)$, $U$ and $V$ being arbitrary functions. (The
arbitrariness in the choice of $\Psi$ reflects the residual coordinate
reparametrization invariance in the conformal gauge. Given $U$ and $V$
correspond to a particular choice of conformal coordinates.)
The two-dimensional metric is
\begin{equation}\begin{array}{lll}
ds^2&=&4[N(\phi)-M_0]\partial_u\Psi\partial_v\Psi dudv\\
&=&4[N(\phi)-M_0]dUdV\,,\label{metric1}
\end{array}\end{equation}
or, using the coordinates [$\phi\equiv\phi(U+V),T\equiv U-V$],
\be
ds^2=-[N(\phi)-M_0]dT^2+[N(\phi)-M_0]^{-1}d\phi^2\,.\label{metric2}
\ee
The general solution depends on the single variable $\phi$. This result is
usually known as the {\it Generalized Birkhoff Theorem} (see e.g.\
\cite{Filippov,LMK,YK,KS}). Finally, substituting (\ref{V}) in Eq.\
(\ref{metric2}) and using (\ref{metric-N}) we have
\be 
ds^2_N=-\left[1-J/R^{N-3}\right]d\tau^2+
\left[1-J/R^{N-3}\right]^{-1}dR^2+R^2d\Omega_{N-2}\,,\label{metric-bh}
\ee
where
\ba
&&\tau=(N-2)\gamma^{-1/(N-2)}t\,,\label{tau}\nonumber\\\nonumber\\
&&R=(\gamma\phi)^{1/(N-2)}\,,\label{R}\\\nonumber\\
&&J={\gamma^{(N-1)/(N-2)}\over (N-2)^2}M_0={\gamma\over N-2}M_{ADM}
\label{J}\,.\nonumber
\ea
Let us conclude this section with a couple of remarks. We have seen that
two-dimensional dilaton gravity can be described by a two-dimensional
nonlinear sigma model with a given target space metric. In particular, for
$N$-dimensional spherically symmetric gravity the fields appearing in the
conformal sigma model are the dilaton and the ADM mass, i.e.\ quantities which
have a direct physical interpretation. The description of spherically
symmetric gravity in terms of geometrical variables is essential for the
quantization of the model since the quantum fields can be directly related to
the original spacetime geometry and problems related to their interpretation
do not show up. The equivalence between the nonlinear sigma model action
(\ref{action-new}) and the dilaton gravity action (\ref{action-dilaton-2}) can
be proved at the canonical level as well. This has been done in
Ref.\ \cite{Cavaglia-PRD}. The general canonical transformation includes, as
particular cases, the canonical transformations discussed in Ref.\
\cite{Varadarajan} for the CGHS \cite{CGHS} model and Ref.\ \cite{Kuchar} for
the four-dimensional black hole.
\section{Sigma model curvature expansion}
The nonlinear sigma model (\ref{action-new}) can be quantized perturbatively
by expanding the target space metric in Riemann normal coordinates \cite{AFM}.
Let us define the adimensional mass ${\cal M}=\gamma^{2/(N-2)} M$. The bulk
term of the action (\ref{action-new}) in the conformal gauge can be cast in
the form [$\sigma\equiv(u,v)$]
\be 
S=\int d^2\sigma\,G_{ij}(X)\partial_\mu X^i\partial^\mu X^j\,,
\label{action-sigma}
\ee
where $X^0\equiv{\cal M}$, $X^1\equiv\phi$ and the metric of the target space
is 
\be
G_{ij}(X)={1\over {\cal N}(X^1)-X^0}\left(\matrix{0&1/2\cr
1/2&0\cr}\right)\,,~~~~~{\cal N}(X^1)=\gamma^{2/(N-2)}
N(\phi)\,.\label{metric-target} \ee
Now we expand the target metric (\ref{metric-target}) in Riemann normal
coordinates around a point $X(0)$ (vacuum expectation value). At the
second order in the Riemann expansion the metric is \cite{Petrov}
\be
G_{ij}(X)=G_{ij}[X(0)]-{1\over 3}R_{ijkl}[X(0)]x^k
x^l-{1\over 3!}R_{ijkl;m}[X(0)]x^k x^l x^m+O(x^4)\label{G-Riemann}
\ee
where $X^i=X^i(0)+x^i$, $x^i\equiv \delta X^i$. Using Eq.\
(\ref{metric-target}) and substituting Eq.\ (\ref{G-Riemann}) in Eq.\
(\ref{action-sigma}) the action at the second order in the Riemann expansion is
\be 
S=\int d^2\sigma\,{\cal L}\,,~~~~{\cal L}={1\over 2}\left[\partial_\mu
y^i\partial^\mu y_i+g\varepsilon_{ij}\varepsilon_{kl}\partial_\mu
y^i\partial^\mu y^k y^j y^l(1+\bar g A_q y^q+O(y^2))\right]\,.
\ee
Here $\varepsilon_{ij}$ is the two-dimensional completely antisymmetric Levi
Civita tensor, $y^i=(x^0\pm x^1)/\sqrt{2}$, $A_q=(1+\tilde g,1-\tilde g)$,
and $g$, $\bar g$, $\tilde g$ are the adimensional coupling constants
\ba
g&=&-\displaystyle{1\over 3}{{\cal V}[X^1(0)]\over {\cal N}[X^1(0)]-X^0(0)}=
-{1\over 3}{V(\phi_0)\over N(\phi_0)- M_0}\,,\label{g}\\\nonumber\\
\bar g&=&\displaystyle{1\over 2\sqrt{2}}{1\over {\cal N}[X^1(0)]-X^1(0)}
=\displaystyle{1\over 2\sqrt{2}}{\gamma^{-2/(N-2)}\over
N(\phi_0)-M_0}\,,\label{barg}\\\nonumber\\ \tilde g&=&\displaystyle\left[{{\cal
V}'[X^1(0)]\over {\cal V}[X^1(0)]}\left[{\cal N}[X^1(0)]-X^0(0)\right]- {\cal
V}[X^1(0)]\right]\left[{\cal N}[X^1(0)]-X^0(0)\right]\nonumber\\\nonumber\\
&=&\displaystyle\left[{V'(\phi_0)\over V(\phi_0)}\left[N(\phi_0)-M_0\right]-
V(\phi_0)\right]\left[N(\phi_0)-M_0\right]\,\gamma^{4/(N-2)}\,,\label{tildeg}
\ea
where ${\cal V}=d{\cal N}/dX^1$. Using Eqs.\ (\ref{V}), (\ref{M-ADM}) and
(\ref{R}) $g$, $\bar g$ and $\tilde g$ read
\ba
g&=&-{\kappa\over 3}{N-3\over N-2}\left({l_{pl}\over
R}\right)^{N-2}\displaystyle{1\over 1-\displaystyle{\kappa\over
N-2}{M_{ADM}\over  m_{pl}}\left({l_{pl}\over
R}\right)^{N-3}}\,,\label{g2}\\\nonumber\\  
\bar g&=&{1\over 2\sqrt{2}}{\kappa^{(N-3)/(N-2)}\over
(N-2)^2}\left({l_{pl}\over R}\right)^{N-3}\displaystyle{1\over
1-\displaystyle{\kappa\over N-2}{M_{ADM}\over  m_{pl}}\left({l_{pl}\over
R}\right)^{N-3}}\,,\label{gbar2}\\\nonumber\\
\tilde g&=&-(N-2)^4\kappa^{-(N-4)/(N-2)}\left({l_{pl}\over R}\right)^{4-N}
\cdot\nonumber\\
&&~~~~~\cdot\left[1-{\kappa\over
(N-2)^2}{M_{ADM}\over 
m_{pl}}\left({l_{pl}\over R}\right)^{N-3}\right]\left[1-{\kappa\over
N-2}{M_{ADM}\over  m_{pl}}\left({l_{pl}\over
R}\right)^{N-3}\right]\,.\label{gtilde2} \ea
where $\kappa=16\pi/V_{N-2}$. Let us investigate the behavior of the coupling
constants. For $R\to\infty$ and fixed $M_{ADM}$, i.e.\ in the asymptotically
flat region far away from a black hole of given (classical) mass $M_{ADM}$,
Eqs.\ (\ref{g2})-(\ref{gtilde2}) read
\ba
g_{\infty}&\sim& -{\kappa\over 3}{N-3\over N-2}\left({l_{pl}\over
R}\right)^{N-2}\left[1+O\left({l_{pl}\over
R}\right)^{N-3}\right]\,,\nonumber\\
\bar g_{\infty}&\sim&{1\over 2\sqrt{2}}{\kappa^{(N-3)/(N-2)}\over
(N-2)^2}\left({l_{pl}\over R}\right)^{N-3}\left[1+O\left({l_{pl}\over
R}\right)^{N-3}\right]\,, \label{largeR} \\
\tilde g_{\infty}&\sim&-(N-2)^4\kappa^{(-N-4)/(N-2)}\left({l_{pl}}\over
R\right)^{4-N}\left[1+O\left({l_{pl}\over R}\right)^{N-3}\right]\,.
\nonumber 
\ea
As expected, the Riemann expansion is an expansion in powers of the curvature,
i.e.\ in powers of $l_{pl}/R$. The theory becomes free in the asymptotically
flat region where the first order correction to the free theory is of order
$O\left((l_{pl}/R)^{N-2}\right)$. The perturbative expansion fails on the
black hole horizon, the coupling constants $g$ and $\bar g$ blowing up when
$R^{N-3}\to J$. The perturbative Riemann expansion is also valid for large
values of $M_{ADM}/m_{pl}$ at distances
\be
R-J^{1/(N-3)}\sim l_{pl}\left({M_{ADM}\over m_{pl}}\right)^{1/(N-3)}\,.
\label{R-J}
\ee
In this regime the dimensional coupling constants (\ref{g2})-(\ref{gtilde2})
read
\ba
&& g \sim \left(\frac{m_{pl}}{M_{ADM}}\right)^{(N-2)/(N-3)}\,, \nonumber \\
&& \bar g \sim \left(\frac{m_{pl}}{M_{ADM}}\right)\,, \label{largeM}\\
&& \tilde g \sim\left(\frac{m_{pl}}{M_{ADM}}\right)^{(4-N)/(N-3)}\nonumber
\ea
and the Riemann expansion is an expansion in powers of $m_{pl}/M_{ADM}$. The
theory becomes a free field theory when $m_{pl}/M_{ADM} \ll 1$. 
\section{Perturbative quantization}
In this section we quantize the theory at one-loop and at the first order in
the curvature expansion ($\bar g=\tilde g=0$). Since the target space is not
Ricci-flat, the conformal symmetry is not preserved at the quantum level.
Conformal symmetry breaking implies running coupling constants and effective
terms in the action that depend on the conformal factor. Since quantum
corrections to the ADM mass due to these terms are subdominant we will postpone
their discussion at the end of the section and work in the unit gauge.  

The vacuum-to-vacuum amplitude is (for notations see Ref.\ \cite{Ramond})
\be
W[J]=N\int {\cal D}[y^i]\,e^{iS[y,J]}\,,~~~~~S[y,J]=\int d^2\sigma\,[{\cal
L}+J_i y^i]\,.
\ee
The free two-points Green function (propagator) is
\ba
<y_i(\sigma_1)y_j(\sigma_2)>&=&i\eta_{ij}\int
{d^2p\over (2\pi)^2}{1\over
p^2+i\epsilon}e^{-ip(\sigma_1-\sigma_2)}\,,\nonumber\\
&=&-\,\eta_{ij}\,{1\over 4\pi}\,\ln{\left[(\sigma_1-\sigma_2)^2\right]}\,. \ea
At the first order in the Riemann expansion the perturbative potential is
\be
V_1(y)={g\over 2}\,\varepsilon_{ij}\,\varepsilon_{kl}\,\partial_\mu
y^i\partial^\mu y^k y^j y^l\,.
\label{V_1}
\ee
The corresponding Feynman rule for the interaction vertex is
\be
\epsffile{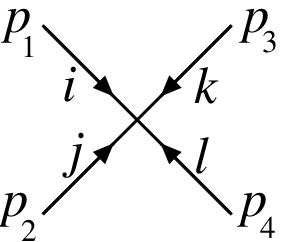}~\hbox{\vbox to
23truemm{\null\vskip 7truemm\hbox{$=-i\displaystyle{g\over
4!}\,(2\pi)^2\delta\left(\sum
p_i\right)[\varepsilon_{ij}\,\varepsilon_{kl}(p_1-p_2)(p_3-p_4)+$}\break
\hbox{$~~~+\varepsilon_{ik}\,\varepsilon_{kl}(p_1-p_3)(p_2-p_4)+
\varepsilon_{il}\,\varepsilon_{kl}(p_1-p_4)(p_3-p_2)]\,.$}\vfill}}
\label{vertex}
\ee
The one-loop correction of the two-point Green function is
\be
\epsffile{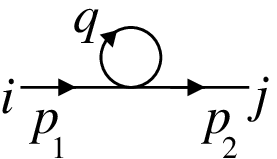}~\hbox{\vbox to
15truemm{\null\vskip 3truemm\hbox{$=\,(2\pi)^2\delta\left(\sum
p_i\right)\displaystyle\prod{1\over p^2_i+i\epsilon}
\Gamma^{(2)}_{ij}\,,$}\vfill}}\label{two-point} \ee
where
\be
\Gamma^{(2)}_{ij}=-g\,\eta_{ij}\int\,{d^2q\over(2\pi)^2}{q^2+p_1^2\over
q^2+i\epsilon}\,.\label{two-point-gamma}
\ee
The (on-shell) one-loop correction of the four-point Green function
($s$-channel) is
\be
\epsffile{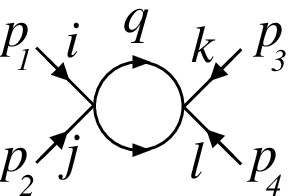}~\hbox{\vbox to
15truemm{\null\vskip 2truemm\hbox{$=\,(2\pi)^2\delta\left(\sum
p_i\right)\displaystyle\prod{1\over
p^2_i+i\epsilon}\Gamma^{(4)}_{ijkl}\,.$}\vfill}}\label{candy}
\ee
where
\be
\Gamma^{(4)}_{ijkl}=g^2\int\,{d^2q\over(2\pi)^2}{1\over(q^2+i\epsilon)
[(p_1+p_2-q)^2+i\epsilon]}(\eta_{ik}\eta_{jl}A_1+\eta_{il}\eta_{jk}A_2+
\eta_{ij}\eta_{kl}A_3)\,,\label{candy-gamma}
\ee
$$\begin{array}{l}
A_1=4(p_1p_2)[(p_1p_2)-q^2+q(p_1+p_2)]+q^2[q^2-2q(p_1+p_2)]-
2[(qp_2)(qp_3)+(qp_1)(qp_4)]\,,\\\\
A_2=4(p_1p_2)[(p_1p_2)-q^2+q(p_1+p_2)]+q^2[q^2-2q(p_1+p_2)]-
2[(qp_1)(qp_3)+(qp_2)(qp_4)]\,,\\\\
A_3=8[q(p_1-p_2)][q(p_3-p_4)]\,.
\end{array}$$
The two- and four-point Green functions (\ref{two-point}) and (\ref{candy})
are infrared and ultraviolet divergent. The infrared divergence can be
eliminated by inserting a infrared regulator. We will check {\em a posteriori}
the consistency of this procedure by proving that the theory is asymptotically
free in the ultraviolet region, i.e.\ that the theory is perturbative for large values of the energy. In
order to regularize the theory we have to compute the ultraviolet
divergences. Using dimensional regularization the divergence of the two-point
Green function (\ref{two-point}) is (details of the calculation
are given in appendix)
\be
[{\rm divergence}~\Gamma^{(2)}_{ij}]\,=i{g\over
2\pi\epsilon}\eta_{ij}p^2\,,\label{div-two-point}
\ee
where $\epsilon=2-d$. 
The divergence above is eliminated by inserting in the Lagrangian density the
counterterm (minimal subtraction)
\be 
{\cal L}^{(2)}={1\over
2}\left(-{g\over 2\pi\epsilon}\right)\partial_\mu
y^i\partial^\mu y_i\,.\label{counterterm-2}
\ee
The divergence of the four-point Green function  (\ref{candy}) is
($s+t+u$-channels)
\be
[{\rm divergence}~\Gamma^{(4)}_{ijkl}]\,=i{11\over
2\pi\epsilon}\mu^\epsilon g^2[\eta_{ij}\eta_{kl}s+\eta_{ik}\eta_{jl}t+
\eta_{il}\eta_{jk}u]\,,\label{div-candy}
\ee
where $s=(p_1+p_2)^2$, $t=(p_1+p_3)^2$ and $u=(p_1+p_4)^2$ are the Mandelstam
variables. The divergence (\ref{div-candy}) is eliminated by inserting in
the Lagrangian density the counterterm
\be
{\cal L}^{(4)}={1\over 2}g\mu^\epsilon\left(-{11\over
6\pi}{g\over\epsilon}\right) \varepsilon_{ij}\,\varepsilon_{kl}\,\partial_\mu
y^i\partial^\mu y^k y^j y^l\,.\label{counterterm-4}
\ee
Finally, the one-loop renormalized Lagrangian density is
\be
{\cal L}_{ren}={1\over 2}Z_1\partial_\mu y^i\partial^\mu
y_i+{1\over 2}g\mu^\epsilon Z_2\varepsilon_{ij}\,\varepsilon_{kl}\,\partial_\mu
y^i\partial^\mu y^k y^j y^l\,,
\ee
where
\be
Z_1=1-{g\over 2\pi\epsilon}\,,~~~~~~Z_2=1-{11g\over 6\pi\epsilon}\,.
\label{Zetas}
\ee
Now we can calculate the $\beta$-function and the anomalous dimension
$\gamma(g)$ of the $y$ fields at one-loop. The result is
\ba
&&\beta(g)=-{5\over 6\pi}g^2+O(g^3)\,,\label{beta}\\
&&\gamma(g)={g\over 2\pi}+O(g^2)\,.\label{anomalous}
\ea
Integrating Eq.\ (\ref{beta}) we obtain
\be
\displaystyle g=g_s{1\over 1+\displaystyle{5\over
6\pi}g_s\ln{\mu\over\mu_s}}\,,\label{g-running}
\ee
where $g_s$ is the value of the coupling constant $g$ at the renormalization
scale $\mu_s$. From Eq.\ (\ref{g-running}) we see that $g\to 0$ for
$\mu\to\infty$, i.e.\ the theory becomes free at high energy scales (asymptotic
freedom). The perturbative regime of the theory is realized at short distances,
where the theory itself exhibits an ultraviolet stable fixed point. Since the
model is asymptotically free in the ultraviolet region, it is possible to
neglect the dependence of the Green functions on the infrared regulator.
Solving the renormalization group equation at one-loop for the $N$-point Green
function, we obtain
\be
<y_1 y_2\dots y_N;g,\mu>=\left({g\over g_s}\right)^{{3\over
10}N}<y_1 y_2\dots y_N;g_s,\mu_s>\,.
\ee
Now let us evaluate the two-point Green function at one loop. We have
\be
<y_i(\sigma_1) y_j(\sigma_2)>=-\eta_{ij}{1\over
4\pi}\left[1-{g\over
4\pi}\ln\left({\mu\over m}\right)^2+O(g^2)\right]
\ln\left[(\sigma_1-\sigma_2)^2\right]\,. 
\ee
In term of the fields $x^i$ the only Green function different from zero is
\be
<x^0(\sigma_1)x^1(\sigma_2)>=-{1\over
4\pi}\left[1-{g\over
4\pi}\ln\left({\mu\over m}\right)^2+O(g^2)\right]
\ln\left[(\sigma_1-\sigma_2)^2\right]\,.
\ee
As a result, at the first order in the curvature expansion, the 
one-loop quantum correction to the Schwarzschild mass is identically zero
\be
<\delta {\cal M}(\sigma_1)\,\delta {\cal M}(\sigma_2)>\,=\,O(g^2)\,.
\ee

Actually, a simple observation shows that the two-point Green function at first
order in the curvature expansion is zero at any loop. This result follows from
the invariance of the interaction Lagrangian density, Eq.\ (\ref{V_1}), under
Poincar\'e group transformations in the $y$ field space. Since the interaction vertex
(\ref{vertex}) has two $\delta {\cal M}$ and two $\delta\phi$ legs, and the
propagator is anti-diagonal in the fields ($\delta {\cal M},\delta\phi$), any
two-point Green function is necessarily diagonal [antidiagonal] in the $y^i$
[$\delta{\cal M},\delta\phi$] fields, respectively.

Let us now consider the perturbative potential at second order in the curvature
expansion 
\be
V_2(y)={1\over 2}g\bar g\varepsilon_{ij}\varepsilon_{kl}A_q\partial_\mu
y^i\partial^\mu y^k y^j y^ly^q\,.
\label{V_2}
\ee
This interaction breaks the Poincar\`e invariance in the $y$ field space. We have
two different vertices:
\be
\begin{array}{lcr}
\epsffile{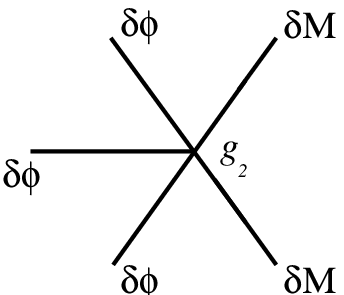}&~~~~~~&\epsffile{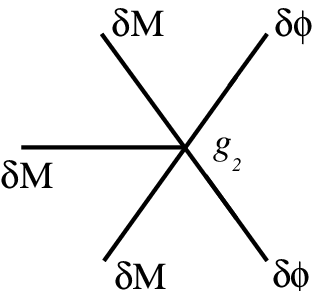}
\label{five-legs}
\end{array}
\ee
where $g_2\sim g\bar g\tilde g$. 

Since $V_2(y)$ is an odd functional of the $y$ fields we expect the interaction
Lagrangian density (\ref{V_2}) to give a non-vanishing two-loop contribution
to the one-point Green function $<\delta{\cal M}>$ 
\be
\begin{array}{c}
\epsffile{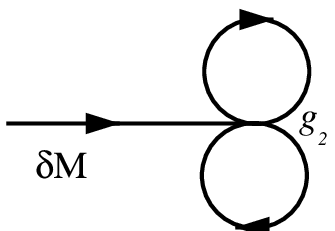}
\label{onep}
\end{array}
\ee
However, a straightforward computation of this diagram shows that $<\delta{\cal
M}>$ is identically zero. The first non-vanishing two-point Green function for
the $\delta M$ field at the second order in the curvature expansion is given by
the Feynman diagram
\be
\begin{array}{c}
\epsffile{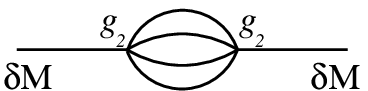}
\label{second-order}
\end{array}
\ee
The above diagram gives a non-diagonal logaritmic divergence to the
propagator of the $y$ fields. By adding the appropriate counterterm in 
the Lagrangian, the first finite, non-zero, two point Green function
for the $\delta {\cal M}$ field is
\be
<\delta {\cal M}(\sigma_1)\,\delta {\cal M}(\sigma_2)>=-\frac{1}{4\pi}\,
A\ln\left(\frac{\mu}{m}\right)\,(1+B\tilde{g}^2)(g\bar{g})^2
\,\ln[(\sigma_1-\sigma_2)^2]\,,
\label{corrM}
\ee 
where $A,B$ are constant adimensional factors.

Using Eq.\ (\ref{corrM}) we can estimate the upper limit of the quantum
corrections of $M_{ADM}$. Since the perturbative expansion fails when
$g\ln(\mu)\sim O(1)$ from Eq.\ (\ref{corrM}) we have 
\be
(\Delta M_{ADM})^2\,\laq\, m_{pl}^2\,|(1+B\tilde{g}^2)\,g\,\bar g^2|\,.
\ee
Recalling Eqs.\ (\ref{largeR})and (\ref{largeM}) we obtain
\be
\Delta M_{ADM}\,\laq\, m_{pl}\,\left(\frac{l_{pl}}{R}\right)^{N/2}\,,
\label{corrM2}
\ee
and
\be
\Delta M_{ADM}\,\laq\, m_{pl}\left(\frac{m_{pl}}{M_{ADM}}\right)^{\frac{N}{2(N-3)}}
\label{corrM3}
\ee
for $R/l_{pl}\gg 1$ and $m_{pl}/M_{ADM}\ll 1$, respectively.

Up to now we have neglected effective terms due to the dependence of the action
on the conformal factor in $d=2-\epsilon$ dimensions. Equations (\ref{corrM2})
and (\ref{corrM3}) make sense only if the contributions due to these terms are
subleading. This is indeed the case, as we have anticipated at the beginning of
this section.  

In the conformal gauge the one-loop renormalized sigma-model action at the
second order in the curvature expansion is
\be
S=\int d^d\sigma\, e^{-\epsilon\Psi/2}\,{\cal L}_{ren}\,+S_{\Psi}+S_{ghost}\,.
\ee
Here $S_{\Psi}$ is the effective action due to the Weyl anomaly and
${\cal L}_{ren}$ is renormalized Lagrangian 
\be
{\cal L}_{ren}={1\over 2}Z_1\partial_\mu y^i\partial^\mu
y_i+{1\over 2}g\mu^\epsilon Z_2\varepsilon_{ij}\,\varepsilon_{kl}\,\partial_\mu
y^i\partial^\mu y^k y^j y^l
+{1\over 2}g\bar{g}\mu^\epsilon\,
Z_3\varepsilon_{ij}\,\varepsilon_{kl}\,\partial_\mu
y^i\partial^\mu y^k y^j y^lA_q\,y^q\,,
\ee
where the divergent part of the renormalization constant $Z_3=1+C g/\epsilon$
comes from the one loop Feynman diagram with one four and one five point
vertices, respectively. Note that the conformal factor $\Psi$ is a propagating
field because the Weyl anomaly action contains a kinetic term for $\Psi$.

Expanding around $d=2$ the leading order $\Psi$-dependent term in the effective
Lagrangian that violates Poincar\'e invariance in the $y$ field space is
(recall that quantum corrections to the ADM mass are only generated by
these terms)
\be
{\cal L}^{CF}_1(\Psi)\sim g^2\bar{g}\,
\varepsilon_{ij}\,\varepsilon_{kl}A_q\,\partial_\mu y^i\partial^\mu y^k y^j y^l
y^q\,\Psi \,.\label{CF}
\ee
Equation (\ref{CF}) gives a contribution to $<\delta{\cal
M}(\sigma_1)\delta{\cal M}(\sigma_2)>$ of order $O(g^4\bar g^2\tilde g^2)$
which is subleading to Eq.\ (\ref{corrM}). Finally, at three loops Eq.\
(\ref{V_2}) originates the (Poincar\'e breaking) term  
\be
{\cal L}^{CF}_2(\Psi)\sim(g\bar{g}\tilde{g})^2\,
\partial_\mu y^i\partial^\mu y^j \Sigma_{ij}\Psi\,,
\label{CF2}
\ee 
where $\Sigma_{ij}$ is a symmetric $2\times 2$ matrix. Equation (\ref{CF2}) 
gives a one-loop contribution to $<\delta{\cal M}(\sigma_1)\delta{\cal
M}(\sigma_2)>$ of order $O\left((g\bar{g}\tilde{g})^4\right)$ whose counterterm
is again subleading to Eq.\ (\ref{corrM}). 
\section{Conclusion}
Let us summarize the main results of the paper. $N$-dimensional spherically
symmetric gravity {\em in vacuo} can be reduced to a two-dimensional conformal
nonlinear sigma model form, Eq.\ (\ref{action-sigma}). The field content of the
latter is given by two fields, $M(\sigma)$ and $\phi(\sigma)$. The $M$ field is
constant on the classical solutions of the field equations. It coincides, apart
from a constant factor, to the ADM mass of the system. The second field,
$\phi$, is related to the radius of the $(N-2)$-dimensional
(hyper)sphere [see Eqs.\ (\ref{metric-bh}) and (\ref{R})]. So both fields have
a direct physical meaning. This property makes the conformal nonlinear sigma
model formulation very attactive. Quantization of the theory in the $(M,\phi)$
representation gives quantum corrections to the mass and to the radius in a
direct and straightforward way. This result cannot be achieved in the
usual Einstein [Eq.\ (\ref{EH})] or dilaton gravity [Eq.\ (\ref
{action-dilaton})] approaches. 

The perturbative quantization of the theory is straightforward. Firstly, the
nonlinear sigma model target space is expanded in Riemann normal coordinates,
i.e.\ in powers of the target space curvature. Then the theory is quantized --
at any order -- by usual quantum field theory techniques. The perturbative
expansion fails on the black hole horizon(s) where the target space metric
exhibits a singularity. This is not surprising: On the horizon(s) strong
quantum gravity effects manifest themselves and a perturbative quantization
must necessarily fail. Conversely, far away from the horizon(s)
quantum gravity effects are weak, the sigma model target space is asymptotically
flat and a perturbative treatment is possible. [The perturbative results hold
also at distances of the order of the horizon(s) for black holes
with large mass in Planck units -- see Eq.\ (\ref{R-J}).]

In this paper we have discussed first and second order corrections in the
curvature expansion. Surprisingly, first order corrections to the ADM mass are
identically zero {\em at any loop}. This follows from the invariance of the
first order interaction under Poincar\`e transformations in the ($\delta
M,\delta\phi)$ field space and from the antidiagonal form of the field
propagator. Therefore, quantum corrections to the ADM mass of a
four-dimensional black hole are not of order $\Delta M_{ADM}\sim R^{-1}$, as
one would naively expect. The first nonzero quantum corrections to $M_{ADM}$
arise (at least) at second order in the curvature expansion [see Eq.\
(\ref{corrM2}) and (\ref{corrM3})].

Equations (\ref{corrM2}) and (\ref{corrM3}) are the main contribution of the
paper. Pure quantum gravity effects make the classical ADM mass of a
spherically symmetric black hole fluctuate according to Eqs.\ (\ref{corrM2})
and (\ref{corrM3}). Hopefully, this result may help to shed light on open
issues in quantum gravity and black hole physics, such as information loss,
black hole thermodynamics and black hole evaporation.
\section*{Acknowledgments}
M.C.\ is partially supported by a Human Capital and Mobility grant of the
European Union, contract number ERBFMRX-CT96-0012, and by a FCT grant Praxis
XXI - Forma{\c c}Ç{\~a}o Avan{\c c}ada de Recursos Humanos, Subprograma
Ci{\^e}ncia e Tecnologia do $2^o$ Quadro Comunit{\'a}rio de Apoio, contract
number BPD/20166/99. We are grateful to Mariano Cadoni and Stanley Deser for
interesting discussions and useful suggestions connected to the subject of this
paper.
\section*{Appendix}
In this appendix we sketch the evaluation of the Feynman diagrams
(\ref{two-point}) and (\ref{candy}). The integrals are evaluated by dimensional
regularization.

\subsection*{Two-point Green function}
By dimensional regularization the calculation of the one-loop two-point Green
function (\ref{two-point}) is reduced to the evaluation of the (Euclidean)
integral in $d=2-\epsilon$ dimensions
\be
I^{(E)}_2=\mu^\epsilon\int{d^{2-\epsilon}q\over (2\pi)^2}{1\over
{q^2+m^2}}\,,\label{I2}
\ee
where we have added a regulator $m^2$ to avoid the infrared divergence at
$q=0$. Indeed, discarding the ultraviolet quadratically divergent integral
(see e.g.\ \cite{Collins}) and Wick rotating to the Euclidean space 
(see Fig. 1, Eq.~(\ref{two-point-gamma})) can be written
\be
\Gamma^{(2)}_{ij}=ig\,\eta_{ij}\,p^2 I^{(E)}_2\,.\label{two-point-gamma2}
\ee
\begin{figure}
\centerline{\epsfxsize=8.0cm
\epsffile{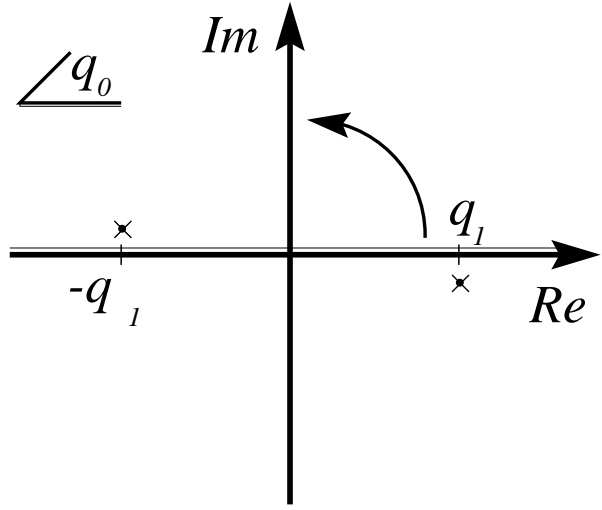}
}
\caption{Wick rotation in the $Q_0$ plane. The two crosses represent the
poles of the integral in the Minkowski space, $I_2$. The latter is equal to
$-iI^{(E)}_2$.} 
\label{fig-one}
\end{figure}
The integral (\ref{I2}) can be immediately evaluated (see \cite{Ramond}, Eq.\
(B.16), p.\ 317). The result is
\be
I^{(E)}_2={1\over 4}\pi^{d/2-2}\Gamma(1-d/2)\left({\mu\over
m}\right)^{2-d}\,.
\ee
Expanding around $d=2$ we obtain
\be
I^{(E)}_2={1\over 4\pi}\left({\mu\over
m}\right)^\epsilon\left[{2\over\epsilon}+\psi(1)+{\epsilon\over
4}\left({\pi^2\over 3}+\psi^2(1)-\psi'(1)\right)+\dots\right]\,,\label{I2-2dim}
\ee
where $\psi$ is the digamma function. Substituting Eq.\ (\ref{I2-2dim}) in
Eq.\ (\ref{two-point-gamma2}) we obtain Eq.\ (\ref{div-two-point}).
\subsection*{Four-point Green function}
With a little bit of algebra Eq.\ (\ref{candy-gamma}) can be cast in the form
[$(s+t+u)$-channels, $d=2-\epsilon$ dimensions]
\be
\Gamma=g^2[\eta_{ij}\eta_{kl}C_1+\eta_{ik}\eta_{jl}C_2+\eta_{il}\eta_{kj}C_3]\,,
\label{candy-gamma2}
\ee
where 
\ba
&&C_1=-{9\over 2}ut[I_4(t)+I_4(u)]+(9u+2t)I_2(t)+(9t+2u)I_2(u)\,,\nonumber\\
&&C_2=-{9\over 2}us[I_4(s)+I_4(u)]+(9u+2s)I_2(s)+(9s+2u)I_2(u)\,,\\
&&C_3=-{9\over 2}st[I_4(s)+I_4(t)]+(9t+2s)I_2(s)+(9s+2t)I_2(t)\,,\nonumber
\ea
and
\be
I_4(p)=\mu^\epsilon\int{d^{2-\epsilon}q\over
(2\pi)^2}{1\over(q^2+i\epsilon)[(p-q)^2+i\epsilon]}\,.\label{I4}
\ee
In the reduction we have discarded the quadratically divergent integrals
\cite{Collins} and used the following relations
$$\begin{array}{l}
\displaystyle\int{d^d q\over (2\pi)^2}{q_i
q_j(d-1)\over(q^2+i\epsilon)[(p-q)^2+i\epsilon]}={p_i p_j\over
2p^2}\left[(d-2)I_2(p)+{dp^2\over 2}I_4(p)\right]+
{\eta_{ij}\over 2}\left[I_2(p)-{p^2\over 2}I_4(p)\right]\,,\\\\
\displaystyle\int{d^d q\over
(2\pi)^2}{p^iq_i\over(q^2+i\epsilon)[(p-q)^2+i\epsilon]}={p^2\over
2}I_4(p)\,,\\\\
\displaystyle\int{d^dq\over
(2\pi)^2}{p^iq_i\over(p-q)^2+i\epsilon}={p^2\over 2}I_2(p)+{1\over
2}\int{d^2q\over (2\pi)^2}{q^2\over(p-q)^2+i\epsilon}\,.
\end{array}$$
$I_4(p)$ can be easily evaluated by inserting an infrared regulator and
performing a Wick rotation similar to the one which is described in Fig.\ 1.
[$I_4(p)$ is infrared divergent and ultraviolet convergent.] Using
\cite{Ramond} [Eq.\ (B.16) p.\ 317] and expanding around $d=2$ we have
\be
I_4(p)=-{i\over 2\pi}\left({\mu\over m}\right)^\epsilon{1\over
p^2\sqrt{1+4m^2/p^2}}\ln{\left|{1+\sqrt{1+4m^2/p^2}\over
1-\sqrt{1+4m^2/p^2}}\right|}\,.\label{I4-2dim}
\ee
Finally, substituting Eq.\ (\ref{I4-2dim}) and Eq.\ (\ref{I2-2dim}) in Eq.\
(\ref{candy-gamma2}) we obtain Eq.\ (\ref{div-candy}).

\end{document}